\documentclass[aps,prd,twocolumn,groupedaddress,nofootinbib]{revtex4-2}
\usepackage{amsmath,amssymb,amsthm,graphicx}
\usepackage[dvipsnames]{xcolor}
\usepackage{multirow}
\usepackage{adjustbox}
\usepackage{comment}
\usepackage{enumerate}
\usepackage{hyperref}
\usepackage{cleveref}
\usepackage{bm}

\newcommand{\Rd}{{\mathord{\mathbb R}^d}}

\newcommand{\grad}{\nabla}

\newcommand{\E}{\mathcal{E}}
\newcommand{\G}{\mathcal{G}}

\newcommand{\R}{\mathcal{R}}

\begin{document}

\title{Which Metric on the Space of Collider Events?}

\author{Tianji Cai}
\affiliation{Department of Physics, University of California, Santa Barbara, CA 93106, USA}
\author{Junyi Cheng}
\affiliation{Department of Physics, University of California, Santa Barbara, CA 93106, USA}
\author{Katy Craig}
\affiliation{Department of Mathematics, University of California, Santa Barbara, CA 93106, USA}
\author{Nathaniel Craig}
\affiliation{Department of Physics, University of California, Santa Barbara, CA 93106, USA}
\affiliation{Physics Division, Lawrence Berkeley National Laboratory, Berkeley, CA 94720, USA}
\affiliation{Berkeley Center for Theoretical Physics, University of California, Berkeley, CA 94720, USA}

\date{\today}

\begin{abstract}
Which is the best metric for the space of collider events? Motivated by the success of the Energy Mover's Distance in characterizing collider events, we explore the larger space of unbalanced optimal transport distances, of which the Energy Mover's Distance is a particular case. Geometric and computational considerations favor an unbalanced optimal transport distance known as the Hellinger-Kantorovich distance, which possesses a Riemannian structure that lends itself to efficient linearization. We develop the particle linearized unbalanced Optimal Transport (pluOT) framework for collider events based on the linearized Hellinger-Kantorovich distance and demonstrate its efficacy in boosted jet tagging. This provides a flexible and computationally efficient  optimal transport framework ideally suited for collider physics applications.
\end{abstract}

\maketitle

\section{Introduction \label{sec:intro}}

The distance between collider events is a promising tool for analyzing particle physics data. But what is the best way to define such a distance? Recently, Komiske {\it et al.} \cite{Komiske:2019fks} pioneered a new approach to the problem using the tools of optimal transport: the {\it Energy Mover's Distance} (EMD)   treats collider events as angular distributions of energy (or closely-related analogues such as transverse momentum distributions in the rapidity-azimuth plane at hadron colliders) and computes a modified form of the canonical Earth Mover's Distance between these distributions. Intuitively, this quantifies the ``work'' required to rearrange one collider event to look like the other, or, in other words, how to ``optimally transport'' one collider event into another.  \emph{Optimal transport} distances  are naturally suited to the space of collider events. They allow for the comparison of raw event data (that is, events composed of a finite number of energy measurements at prescribed locations) without smoothing or binning, and they respect the underlying notion of distance in the reference frame of the detector. Furthermore, they are robust to high frequency noise, ensuring infrared and collinear safety \cite{Komiske:2019fks, Komiske:2020qhg}.
 
 Application of the EMD to collider events has led to new bounds on the modification of infrared- and collider-safe observables by hadronization \cite{Komiske:2019fks}; jet classification using interpretable, distance-based machine learning algorithms \cite{Komiske:2019fks, PhysRevD.102.116019}; visualization of the metric space of jets using CMS Open Data \cite{Komiske:2019jim}; and definition of new shape variables \cite{Cesarotti:2020hwb, Cesarotti:2020ngq}. In \cite{Komiske:2020qhg}, the EMD formed the foundation of a unified approach to collider observables based on the distance between an event and a manifold on the space of events, recasting decades of collider physics in the language of optimal transport. Optimal transport also underlies a number of recently-developed machine learning frameworks for anomaly detection and event generation at the LHC \cite{Stein:2020rou, Kasieczka:2021xcg, Howard:2021pos, DiGuglielmo:2021ide, Kansal:2021cqp, Collins:2021pld}.

The advantages of applying optimal transport to collider events are increasingly clear. But which optimal transport distance is best?  The Earth Mover's Distance, on which the EMD was based, is only one example of a family of balanced optimal transport distances defined between equally normalized distributions, known as {\it $p$-Wasserstein distances}. Although the various $p$-Wasserstein distances are qualitatively similar, they differ in key respects. For instance, only the 2-Wasserstein distance has a Riemannian structure that admits a computationally efficient linearized approximation \cite{lott2006some,AGS}. Additional choices arise when comparing events with unequal total energy. The EMD is obtained by extending the Earth Mover's Distance with an additional term to account for differences in total energy or $p_T$ among collider events, but this extension is far from unique. Rather, there are many possible approaches to the {\it unbalanced} optimal transport problem, one of which enjoys a Riemannian structure akin to the 2-Wasserstein distance. Of course, the ``best'' optimal transport distance for collider events depends on the relevant criteria. Practical considerations include simplicity, robustness, and computational speed, while theoretical considerations (such as the conceptual advantages of a geometric language for collider physics) favor geometric interpretability. 

In this paper, we develop and demonstrate an optimal transport framework for collider events that builds on the EMD with an eye towards these goals. We first situate the EMD in the space of unbalanced optimal transport distances, showing that the EMD is an example of the {\it partial transport distance} \cite{georgiou2008metrics, caffarelli2010free, figalli2010optimal, piccoli2014generalized, piccoli2016properties}. The dynamic formulation of the partial optimal transport distance in turn suggests a general framework for unbalanced optimal transport that may be applied to collider events. In particular, optimizing for both computational speed and geometric interpretability favors an unbalanced optimal transport distance known as the {\it Hellinger-Kantorovich distance} \cite{liero2016optimal,liero2018optimal,chizat2018interpolating,chizat2018scaling,kondratyev2016new}. Loosely speaking, this distance generalizes the 2-Wasserstein distance to unbalanced distributions, preserving a Riemannian structure that lends itself to both geometric interpretation and computational efficiency via linearization \cite{kondratyev2016new,2021arXiv210208807C}. 

The latter advantage is particularly salient. Constructing the geometry of a collider event sample is computationally expensive, requiring the determination of $\mathcal{O}(N_{\rm evt}^2)$ optimal transport distances. However, the computational cost may be significantly decreased by leveraging optimal transport distances admitting a Riemannian structure, such as the 2-Wasserstein distance (for balanced distributions) and the Hellinger-Kantorovich distance (for unbalanced ones). This allows the calculation of optimal transport distances to be linearized by projecting onto the tangent plane at a chosen reference event and computing simpler $\ell^2$ distances on this plane. 

For the balanced case, the Linearized Optimal Transport (LOT) approximation \cite{wang2013linear} to the 2-Wasserstein distance is competitive with EMD in boosted jet classification at a fraction of the computational cost \cite{PhysRevD.102.116019}. Furthermore, as the numerical discretization of the reference event is refined, the linearized metric converges to a true metric on the space of events \cite[Corollary 1]{PhysRevD.102.116019}. Recent work by Delalande and M\'erigot \cite{delalande2021quantitative} has quantified the relationship between the original 2-Wasserstein distance and its linearization, implying, in particular, that LOT preserves  benefits of infrared and collinear safety \cite{Komiske:2019fks, Komiske:2020qhg}.

For the unbalanced case, the corresponding linearization of the Hellinger-Kantorovich distance was recently developed in \cite{2021arXiv210208807C}. In analogy with LOT, we will refer to the linearization of the Hellinger-Kantorovich metric for discrete measures as pluOT (particle linearized unbalanced Optimal Transport). Here we apply pluOT to the classification of boosted $W$ and QCD jets, finding that simple machine learning algorithms based on pluOT perform comparably to the same algorithms based on the EMD in a fraction of the time. This provides a computationally efficient linearized optimal transport framework suitable for comparing collider events with different total energy or $p_T$.

As we will see, the advantages of applying the Hellinger-Kantorovich distance to collider events extend beyond the computational speedup arising from linearization.  The Hellinger-Kantorovich distance and its linearization depend on a length scale $\kappa$ which determines the relative cost of  transporting vs. creating/destroying energy. When applied to the classification of boosted $W$ and QCD jets, simple machine learning algorithms using pluOT with an optimized choice of $\kappa$ outperform the same algorithms using the EMD when the jets are drawn from a sufficiently large $p_T$ range.  We also find that boosted jet classification based on the Hellinger-Kantorovich distance is more robust against pileup contamination than traditional approaches such as $N$-subjettiness.

This paper is organized as follows: In Sec.~\ref{sec:unbalanced} we review the classical Earth Mover's Distance and other $p$-Wasserstein distances of balanced optimal transport. We then turn to the Energy Mover's Distance, showing that it is a special case of the partial transport distance. We use a dynamic formulation of the partial transport distance to introduce a more general family of unbalanced optimal transport distances. Within this family we focus on the Hellinger-Kantorovich distance, reviewing the linearization introduced in \cite{2021arXiv210208807C} and introducing the pluOT framework for collider events. In Sec.~\ref{sec:classification} we consider the classification of boosted $W$ and QCD jets, using the pluOT distance as input to a number of simple machine learning algorithms. We study the performance of these algorithms as a function of the choice of reference measure, Hellinger-Kantorovich scale parameter, and $p_T$ range, finding performance comparable or superior to the same algorithms using EMD distances with much lower computation and storage costs. Finally, in Sec.~\ref{sec:pileup} we study the effects of pileup on pluOT-based classification, finding a surprising level of robustness compared to e.g. classification based on $N$-subjettiness ratios. We conclude in Sec.~\ref{sec:conclusion}, reserving   tables of numerical results for Appendix \ref{app:Tables}.

\section{Unbalanced Optimal Transport} \label{sec:unbalanced}

Consider two discrete measures $\E  , \E' $, which assign positive masses  $\{E_i\}_{i \in I}$, $\{E_j \}_{j \in J}$ to particles at locations $\{ x_i \}_{i \in I}$, $\{x_j'\}_{j \in J}$ in a domain $\Omega \subseteq \Rd$. For example, in the collider physics context, such measures are often used to represent jet events, where the mass assigned at a point represents the energy measured by the calorimeters.\footnote{In what follows we will largely use the lexicon of optimal transport, emphasizing that energy or $p_T$ (rather than jet or particle masses) will play the role of ``mass'' in collider physics applications.} When the measures have the same total mass, $\sum_i E_i = \sum_j E_j'$, the theory of (balanced) optimal transport  provides a natural notion of distance between the two measures. For example, if $d_{ij} = \|x_i-x_j'\|$ represents the distance between particles in each discrete measure, the classical Earth Mover's Distance is given by
\begin{align*}
W_1( \E, \E') &:= \min_{\gamma_{ij} \in \Gamma_{(\E,\E')}^{\rm EMD} }  \sum_{ij}  d_{ij} \gamma_{ij}   \\
\Gamma_{(\E,\E')}^{\rm EMD} &:= \left\{ \gamma_{ij} : \gamma_{ij} \geq 0, \ \sum_j \gamma_{ij} =  {E}_i, \  \sum_i \gamma_{ij} = E_j'   \right\} .
\end{align*}
This can be interpreted as finding a way to rearrange the distribution of mass in $\E$ to match $\E'$, using the least amount of effort: $\gamma_{ij}$  represents the amount of mass moved from particle $i$ to particle $j$ and $\sum_{ij} \gamma_{ij}  d_{ij}$ represents the required effort. By making a mild modification to the notion of effort, one may also consider the $p$-Wasserstein distance between two discrete measures of equal total energy, for $p \geq 1$,
\begin{align}  \label{Wpdef}
 {  W_p}( \E, \E') &:= \min_{\gamma_{ij} \in \Gamma_{(\E,\E')} }  \left( \sum_{ij} \gamma_{ij}  d_{ij}^p \right)^{1/p}  .
\end{align}

In recent years, there has been substantial interest in generalizing the above metric to measures with unequal total energy, known as the \emph{unbalanced} optimal transport problem \cite{caffarelli2010free, figalli2010optimal, piccoli2014generalized, piccoli2016properties, liero2016optimal,liero2018optimal,chizat2018interpolating,chizat2018scaling,kondratyev2016new,gangbo2019unnormalized,ryu2018unbalanced}.
We begin by recalling the notion of \emph{partial optimal transport}, as   already   used in collider physics applications by Komiske, et. al. \cite{Komiske:2019fks}. Next, we describe how partial optimal transport distances relate to the  \emph{Hellinger-Kantorovich} distance \cite{liero2016optimal,liero2018optimal,chizat2018interpolating,chizat2018scaling,kondratyev2016new}, which will be the key tool in the present work. As we will describe, the \emph{Hellinger-Kantorovich} distance is unique among related distances due to the fact that it has a Riemannian structure \cite{kondratyev2016new,2021arXiv210208807C}. This   allows us to linearize the metric and vastly improve its computational performance on classification tasks.

\subsection{Partial Optimal Transport}

The first unbalanced optimal transport metric considered in collider physics was the \emph{Energy Mover's Distance} studied by  Komiske, et. al  \cite{Komiske:2019fks}. In this case,  for fixed $R\geq \max_{ij} d_{ij} /2$,  the distance between   discrete measures $\E, \E'$ is  
\begin{align*}
  {\rm EMD}^R (\E, \E') &= \min_{\gamma_{ij} \in {\Gamma}_{\leq(\E,\E')}^{\rm EMD}} \frac{1}{R} \sum_{ij}d_{ij}  \gamma_{ij} + \left| \sum_i E_i - \sum_j E_j' \right|  ,
\end{align*}
  where a transport plan $\gamma_{ij}$ belongs to the set ${\Gamma}_{\leq(\E,\E')}^{\rm EMD}$ in case it satisfies the following four criteria:
  \begin{enumerate}[(a)]   \itemsep0em 
\item $ \gamma_{ij} \geq 0$,
\item $\sum_j \gamma_{ij} \leq E_i$, \label{econstraint}
\item $\sum_i \gamma_{ij} \leq E_j'$, \label{eprimeconstraint}
\item $  \sum_{ij} \gamma_{ij} = \min \left(\sum_i E_i, \sum_j E_j' \right)$. \label{minenergyconstraint}
\end{enumerate}
 These criteria ensure  that (1) the amount of mass moved between any two particles is always nonnegative, (2) the maximum amount of mass that can be moved from location $i$ in $\E$ to any   location in $\E'$ is $E_i$, (3) the maximum amount of mass that can be moved to location $j$ in $\E'$ from any   location in $\E$ is $E_j'$, and (4) the total mass that is moved equals the total mass of whichever event has  smaller mass. If strict inequality holds in (\ref{econstraint}), we will say $E_i - \sum_j \gamma_{ij}$ mass has been \emph{destroyed} at $x_i$, and if strict inequality holds in (\ref{eprimeconstraint}), we will say $E_j' - \sum_i \gamma_{ij}$ mass has been \emph{created} at $x_j'$.
 Note that, when the measures $\mathcal{E}$ and $\mathcal{E}'$ have equal total mass, ${\rm EMD}^R(\E,\E')  = \frac{1}{R} W_1(\E,\E')$.

In fact, the Energy Mover's Distance is a special case of the \emph{partial  transport distance} studied by Georgiou, Karlsson, and Takyar \cite{georgiou2008metrics}, Caffarelli and McCann \cite{caffarelli2010free}, Figalli \cite{figalli2010optimal}, and Piccoli and Rossi \cite{piccoli2014generalized, piccoli2016properties}:    for $\kappa>0$, $p \geq 1$, define
\begin{align}
  &T_p^\kappa(\E, \E') \nonumber \\
  &=   \min_{\gamma_{ij} \in \Gamma_{\leq(\E,\E')}}   \left(  \Sigma_{ij} d_{ij}^p  \gamma_{ij}  \right)^{1/p} \nonumber \\
    & \quad   \qquad  +   \frac{\kappa}{2}  ( \left| \Sigma_i E_i -  \Sigma_{ij} \gamma_{ij}\right|  +   \left| \Sigma_j E_j' - \Sigma_{ij} \gamma_{ij} \right| )   , \label{Tpkappadiscrete}
\end{align}
where a transport plan $\gamma_{ij}$ belongs to the set $\Gamma_{\leq(\E,\E')}$ in case it satisfies criteria (a,b,c) above.  
The two main differences between ${\rm EMD}^{R}$ and $T_p^\kappa$ are that, first,  the partial transport distances allow $p \geq 1$ and, second, they  permit the amount of mass that is rearranged from $\E$ to $\E'$ to differ from the total mass of whichever event has smaller mass. To see this, assume without loss of generality that $\E$ has smaller total mass, $\sum_i E_i \leq \sum_j E_j'$. The distance ${\rm EMD}^{R}$ requires that all of the mass in $\E$ be rearranged: exactly $\sum_j E_j' - \sum_i E_i $ mass is created in $\E'$, and no mass is destroyed. On the other hand, $T^\kappa_p$ allows for  $\sum_{ij} \gamma_{ij} \in (0, \min (\sum_i E_i, \sum_j E_j' ))$ mass to be rearranged: $\sum_i E_i - \sum_{ij} \gamma_{ij}$ mass is destroyed in $\E$, and $ \sum_j E_j' - \sum_{ij} \gamma_{ij}$ mass is created in $\E'$. 

Now we show why,  for  $\kappa = 2R \geq \max_{ij} d_{ij} $,  ${\rm EMD}^R$ coincides with (a constant multiple of) $T^\kappa_1$. First, note that the EMD constraint set is a subset of the Piccoli-Rossi constraint set,  $ \Gamma^{\rm EMD}_{\leq(\E,\E')} \subseteq \Gamma_{\leq(\E,\E')}$. Furthermore, if $\gamma_{ij} \in \Gamma^{\rm EMD}_{\leq(\E,\E')}$, then the values of the objective function  in each minimization problem coincide, up to a factor of $\kappa = 2R$. Thus, if we can show that $\kappa = 2R \geq \max_{ij} d_{ij} $ ensures that the optimizer $\gamma_{ij}^*$ of $T_1^\kappa$ belongs to the stricter constraint set $\Gamma^{\rm EMD}_{\leq(\E,\E')}$, we can     conclude that
\begin{align} \label{firsttkinequality}
T^\kappa_1(\E, \E') = T^{2R}_1(\E, \E') = R \ {\rm EMD}^R(\E,\E') .
\end{align}
 
Observe that, using properties (\ref{econstraint},\ref{eprimeconstraint}) of the constraint set $\Gamma_{\leq(\E,\E')}$, we may remove the absolute value signs in the definition of $T^\kappa_1$ and express it  equivalently as
\begin{align} &T^\kappa_1(\E, \E') = \nonumber \\
& \quad \min_{\gamma_{ij} \in \Gamma_{\leq(\E,\E')}} \sum_{ij}  \left(   d_{ij} -  \kappa  \right) \gamma_{ij} + \frac{\kappa}{2}  \left(\Sigma_i E_i + \Sigma_j E_j' \right)   \label{altform}
\end{align}
Thus, if $\kappa  \geq \max_{ij}d_{ij}$,  the coefficient on $\gamma_{ij}$ is always negative, so the optimal $\gamma_{ij}^*$ for the $T^\kappa_1$ distance will be as large as possible, subject to the constraints (\ref{econstraint},\ref{eprimeconstraint}). In particular, the optimal $\gamma_{ij}^*$ will satisfy constraint (\ref{minenergyconstraint}) and belong to  $\Gamma^{\rm EMD}_{\leq (\E,\E')}$.

The above argument not only establishes the equivalence between $T^\kappa_1$ and ${\rm EMD}^R$ for $\kappa = 2R \geq \max_{ij} d_{ij} $, but also sheds light on the role of the parameter $\kappa>0$. From (\ref{Tpkappadiscrete}) we observe that smaller $\kappa$ makes creation and destruction cheaper and transport comparatively more expensive. In fact, using (\ref{altform}), we can make this quantitative:  if $\gamma_{ij}^*$ is  the optimizer, then for any $i,j$ such that $d_{ij} >    \kappa $, we must have  $\gamma_{ij}^* = 0$. (If not, we could find a strictly better choice of $\gamma$ in $\Gamma_{\leq(\E,\E')}$ by setting $\gamma_{ij} = 0$, contradicting that $\gamma_{ij}^*$ was the optimizer.) In other words, energy will never be transported over a distance greater than $ \kappa$.

\subsection{From Partial Optimal Transport to the Hellinger-Kantorovich Distance}

One of the key contributions of Piccoli and Rossi's work on the partial optimal transport distance $T^\kappa_p$ is a \emph{dynamic} formulation of the  distance \cite{piccoli2016properties}. This dynamic perspective is most clear when  $T^\kappa_p$ is stated in full generality, as a distance on the space of finite Borel measures $\mathcal{M}(\Omega)$: for $\mu, \mu' \in \mathcal{M}(\Omega)$, $\kappa >0$, and $p \geq 1$,
\begin{align}
&T^\kappa_p(\mu,\mu') \nonumber \\
&= \inf_{\gamma \in \Gamma_{\leq(\mu,\mu')}} \left( \iint |x - x'|^p d \gamma(x,x') \right)^{1/p}  \nonumber \\
&\qquad+ \frac{\kappa}{2} \left( \left| \int \mu - \iint \gamma \right| + \left| \int \mu' - \iint \gamma \right| \right) , \label{Tkappapcontinuum}
\end{align}
where  we say $\gamma \in \Gamma_{\leq(\mu,\mu')}$ in case  $\gamma \in \mathcal{M}(\Omega\times \Omega)$  satisfies $\gamma(B \times \Omega) \leq \mu(B)$ and $\gamma(\Omega \times B) \leq \mu'(B)$ for any Borel set $B$. 
Note that equation (\ref{Tkappapcontinuum}) reduces to equation (\ref{Tpkappadiscrete}) when $\mu = \sum_{i \in I} \delta_{x_i} E_i$ and $\mu' = \sum_{j \in J} \delta_{x_j'}E_j'$. 

Piccoli and Rossi \cite{chizat2018interpolating,piccoli2016properties} show that $T^\kappa$ has the following equivalent dynamic formulation,
\begin{align*}
T^\kappa_p(\mu,\mu') &=  \inf_{\rho, v, \psi \in \mathcal{C}(\mu,\mu')} (A^\kappa_p[\rho,v,\psi])^{1/p} , \\
A^\kappa_p[\rho,v,\psi] &=  \int_0^1   \int_\Omega (|v(x,t)|^p + (\kappa/2) |\psi(x,t)| )\rho(x,t) dx   dt , \\
\mathcal{C}(\mu, \mu') \quad \\
 = \left\{\rho  \in \right. & \left. C([0,1], \mathcal{M}(\Omega)), v \in L^2(d\rho_t dt), \psi \in L^1(d\rho_t dt): \right. \\
&    \left. \partial_t \rho + \nabla \cdot (\rho v) = \psi \rho , \rho(\cdot, 0) = \mu, \ \rho(\cdot, 1) = \mu' \right\} ,
\end{align*}
In other words, one can find the $T^\kappa_p$ distance from $\mu$ to $\mu'$ by considering all curves $\rho$ connecting $\mu$ to $\mu'$ with velocity $v$ and reaction rate $\psi$ and finding the curve with least action $A^\kappa_p[\rho,v, \psi]$.

This dynamic perspective  reveals a   general framework for unbalanced optimal transport problems, in terms of minimizing different notions of action. In particular, as observed in \cite{chizat2018interpolating}, for any $\kappa >0$, $p \geq 1$, and $q \geq 1$, one may consider
\begin{align*}
&A^\kappa_{p,q}[\rho,v,\psi] \\
&\quad \quad =  \int_0^1   \int_\Omega (|v(x,t)|^p + (\kappa/2)^{q} |\psi(x,t)|^q )\rho(x,t) dx   dt .
\end{align*}

As before, large values of $\kappa >0$ penalize creation and destruction. In particular, sending $\kappa \to +\infty$ \cite[Theorem 7.24]{liero2018optimal}, 
\begin{align*}
\lim_{\kappa \to +\infty} \inf_{\rho, v, \psi \in \mathcal{C}(\mu,\mu')} & \left(  A^\kappa_{p,q}[\rho,v,\psi] \right)^{1/p} \\
& \qquad \qquad =
\begin{cases} W_p(\mu,\mu') &\text{ if }\int \mu = \int \mu' \\ +\infty &\text{ otherwise.} \end{cases}
\end{align*}

While minimizing the action $A^\kappa_{p,q}[\rho,v,\psi]$ with $q=1$ yields the \emph{partial transport distance} $T^\kappa_p$ described in the previous section,  minimizing it for $p=q=2$ yields the \emph{Hellinger-Kantorovich distance},
\begin{align} \label{HKdynamic}
{\rm HK}^\kappa(\mu,\mu') =  \inf_{\rho, v, \psi \in \mathcal{C}(\mu,\mu')} (A^\kappa_{2,2}[\rho,v, \psi])^{1/2} .
\end{align}
This case is distinguished among all $p, q \geq 1$, since it is the only choice that directly gives rise to an infinite dimensional Riemannian manifold \cite{kondratyev2016new,chizat2018interpolating}. Furthermore, not only does the Hellinger-Kantorovich metric have a well defined limit as $\kappa \to +\infty$ whenever $\mu$ and $\mu'$ have equal mass, $\lim_{\kappa \to +\infty} HK^\kappa(\mu,\mu') = W_2(\mu,\mu')$, the $\kappa \to 0$ limit is also well defined for arbitrary $\mu, \mu'$,
\begin{align} \label{Hellinger} \lim_{\kappa \to 0} \frac{1}{\kappa} HK^\kappa(\mu,\mu') = \left( \int \left| \sqrt{\frac{d\mu}{dx}} - \sqrt{\frac{d\mu'}{dx}} \right|^2 dx \right)^{1/2}, 
\end{align}
 which is known as the \emph{Hellinger distance} \cite{liero2016optimal,liero2018optimal,chizat2018interpolating}.

 Like the original definition of the partial optimal transport distances $T^\kappa_{p}$, it can also be expressed in terms of a static minimization problem, which for simplicity, we state in the case of fully discrete measures $\E, \E'$,
\begin{align}
&{\rm HK}^\kappa(\E, \E') \label{HKKantorovich} \\
& = \min_{\gamma_{ij} \geq 0} \sum_{ij} \left( \ell^\kappa(d_{ij}) \gamma_{ij} +  \kappa^2  {\rm KL}(\G,\E) +  \kappa^2 {\rm KL}(\G',\E') \right)^{1/2}  \nonumber , 
\end{align}
where $\G$ and $\G'$ are auxiliary discrete measures, with $\G$ assigning mass $G_i =  \sum_j \gamma_{ij}$ to location $x_i$ and $\G'$ assigning mass $G_j' = \sum_i \gamma_{ij}$ to $x_j$, and
\begin{align} 
&\ell^\kappa(s) = \begin{cases} -  2 \kappa^2 \log( \cos^2( s/\kappa)) &\text{ if } s < \frac{  \pi}{2}\kappa ,  \\
+\infty, &\text{ otherwise,} \end{cases} \label{lkdef} \\ 
&{\rm KL}(\G,\E) = \sum_i E_i  f \left( \frac{G_i }{E_i} \right) , \quad f(s) = s \log(s)-s+1 . \nonumber
\end{align}
The equivalence between (\ref{HKdynamic}) and (\ref{HKKantorovich}) is a significant mathematical  result, due to Liero, Mielke, and Savar\'e, based on a surprising connection with cone geometry \cite{liero2016optimal,liero2018optimal}.

The optimizer $\gamma_{ij}$ of (\ref{HKKantorovich}) represents how much mass is transported from $x_i$ in $\G$ to $x_j'$ in $\G'$, that is, $\gamma_{ij}$ is the optimal transport plan from $\G$ to $\G'$. In general, $G_i \neq E_i$ and $G_j' \neq E_j'$, and the energy that is not transported can be thought of as having been created or destroyed. In particular,
\begin{itemize}
\item if $G_i > E_i$, we say energy was \emph{created} at $x_i$;
\item if $G_i < E_i$, we say energy was \emph{destroyed} at $x_i$;
\item if $G_j > E_j'$, we say energy was \emph{destroyed} at $x_j$;
\item if $G_j < E_j'$, we say energy was \emph{created} at $x_j$.
\end{itemize}
(Note that the first and third options did not arise for the $T^\kappa_p$ distance, due to requirements (\ref{econstraint}-\ref{eprimeconstraint}) for the set of transport plans $\Gamma_{\leq(\E,\E')}$.) While until now we have always assumed that our discrete measures have strictly positive energy at every location, $E_i, E_j' >0$, observe that it is possible for $G_i$ or $G_j'$ to be zero.

The first term in the minimization problem in equation (\ref{HKKantorovich}) penalizes transporting energy over long distances. As with $T^\kappa_p$, small values of $\kappa$ penalize transport:  energy will never be transported over distance greater than $\frac{\kappa\pi}{2}$ . The second two terms penalize the difference between $\G$ and $\E$ and between $\G'$ and $\E'$, in terms of the Kullback-Liebler divergence.

A major difference between the Hellinger-Kantorovich metric and the 2-Wasserstein metric considered in the authors' previous work  \cite{PhysRevD.102.116019} is that the Hellinger-Kantorovich metric allows for the comparison of events with unequal total energy. However,   even when the total energy of events $\E$ and $\E'$ coincide, ${\rm HK}^\kappa(\E,\E')$ is in general not equal to $W_2(\E, \E')$. This can be seen, for example, from   equations (\ref{HKKantorovich}-\ref{lkdef}): mass will never be transported more than distance $\frac{\kappa \pi}{2}$. Interestingly, the converse is also true: 
if mass is not transported from $x_i$ to $x_j'$, that is, if $\gamma_{ij} = 0$, then we must have $d_{ij}=\|x_i - x_j\| \geq \frac{ \kappa \pi}{2}$ \cite[Lemma 3.13]{2021arXiv210208807C}. 

As already observed in the collider physics context for the special case $\kappa = +\infty$, $p=q=1$ \cite{Komiske:2019fks}, Chizat, et. al, identified that minimizing the action $A^\kappa_{p,q}$  has a dual characterization in terms of the following maximization problem
\begin{align*}
&\inf_{\rho,v, \psi \in \mathcal{C}(\mu,\mu')} A^\kappa_{p,q}[\rho,v,\psi] \\
&\qquad \qquad = \sup_{\varphi \in {\rm HJ}_{p,q}^\kappa} \int_\Omega \varphi(x,1) d \mu'(x) - \int_\Omega \varphi(x,0) d \mu(x) \\
&{\rm HJ}^\kappa_{p,q} = \left\{ \varphi \in  C^1([0,1]\times \Omega)  : \right. \\
&\qquad \qquad \left. \partial_t \varphi + \frac{ |\nabla \varphi|^{p'}}{p'} + (\kappa/2)^{ -q'} \frac{ |\varphi|^{q'}}{q'} \leq 0 \right\} ,
\end{align*}
where $p' = \frac{p}{p-1}$ and $q' = \frac{q}{q-1}$ \cite{chizat2018interpolating}; this has been rigorously justified  in the cases $p=q=2$ \cite{liero2018optimal} and $p=q=1$ \cite{piccoli2016properties}. In the above equations, we use the following convention if either $p =1$ or $q=1$:  as $p \to 1$ or $q \to 1$, the second and third terms in the sum would diverge to $+\infty$ unless $|\grad \varphi| \leq 1$ or $|\varphi| \leq \kappa/2$, in which case each term would converge to zero. Consequently, if either $p=1$ or $q=1$, we drop the respective term from the inequality and add the constraint $|\grad \varphi| \leq 1$ or $|\varphi| \leq \kappa/2$. Furthermore, if we first send $\kappa \to +\infty$, to reduce to the $p$-Wasserstein, since $q' \geq 1$, the second term in the inequality constraint will vanish. In this way, for the case $\kappa =+\infty$, $p=q=1$ considered by Komiske, et. al. \cite{Komiske:2019fks}, we see that we may replace ${\rm HJ}^{+\infty}_{1,1}$ by $\{\phi(x,t)  : \phi(x,t) = \Phi(x) \in C^1(\Rd),  \sup_x | \nabla \Phi(x) | \leq 1 \}$, the space of 1-Lipschitz functions. Indeed, it is the comparatively simple formulation of the dual problem in the case $p=q=1$ for all $\kappa >0$ that makes such metrics popular in practice. However, we choose to work with $p=q=2$, due to its superior geometric properties, which allow us to linearize the metric, vastly improving computational efficiency of our  method, while preserving key features of the optimal transport distance.

\subsection{Linearized Hellinger-Kantorovich Metric} \label{LHKMsection}
We now describe the linearization of the Hellinger-Kantorovich metric, as introduced in \cite{2021arXiv210208807C}, that we use in the present work. We begin by explaining how to construct an embedding of events $\E$ into Euclidean space $\mathbb{R}^{dn} \times \mathbb{R}^{n}$, using the optimal  transport plan $\gamma_{ij}$ in the Hellinger-Kantorovich metric; see equation (\ref{HKKantorovich}). We will then describe how computing the distance between the embeddings provides a linearization of the Hellinger-Kantorovich distance.

Let $\mathcal{R}$ denote a discrete reference measure, consisting of particles at locations $\{x_i\}_{i \in I}$ with positive masses $\{R_i\}_{i \in I}$. For any discrete measure $\E'$, let $\gamma_{ij}$ denote an optimizer of (\ref{HKKantorovich}), which represents an optimal transport plan from the auxiliary measures $\G$ to $\G'$. (Note that there may be more than one optimizer.) In general, the transport plan $\gamma_{ij}$ may send mass from $x_i$ in $\G$ to many different locations in $\G'$. In order to linearize the Hellinger-Kantorovich metric, we first consider the average of these locations, weighted by how much mass is sent to each place and normalized by the amount of mass starting at $x_i$ in $\G$,
\begin{align} \label{approxtranspmap}
z_i = \begin{cases} \frac{1}{G_i} \sum_{j} \gamma_{ij} x_j'   &\text{ if } G_i > 0 \\
x_i &\text{ if }G_i = 0 \end{cases}  
\end{align}

Next, we consider the average amount that mass starting at location $x_i$ needs to be rescaled, via creation or destruction, in order for $\R$ to become $\E'$: for each $x_j'$,    consider the ratio $E_j'/G_j'$, between the amount of mass that must end up at location $x_j'$ and the amount of mass transported by $\gamma_{ij}$ to $x_j'$. If $E_j'/G_j' >1$, mass needs to be created at $x_j$, and if $E_j'/G_j'< 1$, mass needs to be destroyed at $x_j$. Note that this quantity is well-defined only for $G_j' = \sum_i \gamma_{ij} >0$. In fact, this is a necessary assumption for the Hellinger-Kantorovich metric to be linearized in a manner that admits a Euclidean embedding   \cite[p18]{2021arXiv210208807C}. Recall from the previous section that a sufficient condition for $\gamma_{ij} >0$ is  $d_{ij} = \|x_i - x_j'\|< \frac{\kappa \pi}{2}$. Consequently, in what follows, we will suppose that $\kappa$ is sufficiently large so that, for each $x_j'$, there exists $x_i$ so that $\|x_i - x_j'\| < \frac{\kappa \pi}{2}$. This will ensure $G_j'>0$ for all $j$.

With this assumption in hand, we now consider, for each fixed $x_i$,  the weighted average of this ratio, representing how much mass needs to be created/destroyed at $x_j$, with respect to how much mass $\gamma_{ij}$ transports to each $x_j'$, normalized by the amount of mass $G_i$ originally starting at $x_i$,
\begin{align} \label{approxcreationdestruction}
u_i = \begin{cases} \frac{1}{G_i} \sum_j \left(\frac{E_j'}{G_j'} \right) \gamma_{ij}   &\text{ if } G_i >0 , \\
0& \text{ if } G_i = 0 . 
\end{cases}
\end{align}
While the coordinate $z_i$, defined in equation (\ref{approxtranspmap}), represents the average location that mass starting at $x_i$ is transported to in $\E'$, the coordinate $u_i$ represents the average amount of creation/destruction that will happen to mass that started at $x_i$, after it is transported.

With these quantities in hand, we may now state the formula for the linearized Hellinger-Kantorovich metric. In analogy with the linearized 2-Wasserstein metric, which is known as  LOT (Linearized Optimal Transport) \cite{wang2013linear,PhysRevD.102.116019}, we will refer to the linearization of the Hellinger-Kantorovich metric for discrete measures as pluOT (particle linearized unbalanced Optimal Transport), emphasizing that it is a discrete particle approximation of the  continuum linearization of the Hellinger-Kantorovich metric studied in previous work by the first two authors \cite{2021arXiv210208807C},
\begin{align} \label{pluotdef}
{\rm pluOT}^\kappa(\E', \tilde{\E}') \hspace{-2cm}&  \\
\qquad  &= \left( \sum_i R_i   \|v_i - \tilde{v}_i \|^2 +  \frac{\kappa^2}{4}  R_i  |\alpha_i - \tilde{\alpha}_i|^2   \right)^{1/2} , \nonumber \\
v_i &= \kappa   \ {\rm sgn}(z_i-x_i) \sqrt{  u_i G_i/ R_i }\sin(\|z_i-x_i\|/\kappa) , \nonumber \\
\alpha_i &= 2 \left(\sqrt{ u_i G_i/R_i } \cos(\|z_i-x_i\|/\kappa) - 1 \right) \nonumber .
\end{align}
Note that this approximation depends on the choice of the optimal transport plans $\gamma_{ij}, \tilde{\gamma}_{ij}$ via their dependence on $x_i, z_i, \tilde{x}_i, \tilde{z}_i$; see \cref{approxtranspmap,approxcreationdestruction}.  As before, the unusual expressions for $v_i$ and $\alpha_i$ derive from the surprising connection to cone geometry \cite{liero2016optimal,liero2018optimal,chizat2018interpolating}.

In analogy with LOT for balanced optimal transport, a key benefit of the pluOT approximation of the Hellinger-Kantorovich metric is that it provides a natural embedding 
\begin{align} \label{Eucemb}
\E' \mapsto (v_i, \alpha_i)_{i \in I} \in \mathbb{R}^{dn} \times \mathbb{R}^n ,
\end{align}
where $d$ is the dimension of the underlying domain $\Omega$ in which particles are located and $n$ is the number of particles in the discrete reference measure, $n = |I|$. This vector may be interpreted geometrically as an approximation of the tangent vector from $\R$ to $\E$ with respect to the Hellinger-Kantorovich geometry, an interpretation that may be made precise when $\R$ is a finite Borel measure that is absolutely continuous with respect to Lebesgue mesaure \cite[Definition 4.5]{2021arXiv210208807C}. In this way, it is natural to compare two discrete measures $\E'$ and $\tilde{\E}'$ by computing the distance between the vectors $(v_i,\alpha_i)$ and $(\tilde{v}_i,\tilde{\alpha}_i)$ as elements of the tangent space at $\R$, as in  \cref{pluotdef} above.

The above Euclidean embedding, \cref{Eucemb}, is useful from the perspective of  classification algorithms for two reasons. First, while computing the embedding $\E' \mapsto (v_i, \alpha_i)$  for each event $\E'$ in a sample requires  $O(N_{evt})$ computations of the Hellinger-Kantorovich metric in \cref{HKKantorovich}, computing the linearized distance between all pairs of events $\E'$ and $\tilde{\E}'$ using pluOT only requires $O(N_{evt}^2)$ computations of the weighted Euclidean metric; see  \cref{pluotdef}. Given that computing a weighted Euclidean metric is several orders of magnitude faster than computing the Hellinger-Kantorovich metric, our approach using pluOT offers a substantial computational advantage compared to computing the exact Hellinger-Kantorovich distance between all pairs of events.

The second reason that the Euclidean embedding in pluOT is useful in classification tasks is that it allows us to apply a wider range of classification algorithms directly to the vectors $(v_i, \alpha_i)$ and $(\tilde{v}_i, \tilde{\alpha}_i)$ representing the discrete measures $(\E', \tilde{\E}')$, including classification algorithms that require a Euclidean structure. In particular, we are often able to delegate the computation of the entire pairwise distance matrix to efficient downstream methods, leading to a large storage advantage over other methods.

In our analysis of the linearized Hellinger-Kantorovich metric as a tool to classify jets, we will investigate  the effects of creation/destruction in the Hellinger-Kantorovich metric separately from the fact that it allows for the comparison of events with unequal total energy. We do this by separately analyzing the classification performance of the linearization of ${\rm pluOT}^\kappa(\E, \E')$ with the classification performance of  ${\rm pluOT}^\kappa \left( {\E}/{\sum_i E_i}, {\E'}/{\sum_j E_j'} \right) $
where $\E/(\sum_i E_i)$ denotes the normalized measure, in which the mass $E_i$ of each particle in $\E$ is replaced by $E_i /(\sum_i E_i)$. The Hellinger-Kantorovich metric exhibits a simple scaling under this transformation \cite[Theorem 3.3]{laschos2019geometric}: denoting $m = \sum_i E_i$, 
\begin{align} \label{normalizedHKcoords}
(E_j')^{norm} &= m^{-1/2} E_j'  & \gamma_{ij}^{norm} &= m^{-1/2} \gamma_{ij}  , \\
  G_i^{norm} &=  m^{-1/2} G_i &  (G_j')^{norm} &= m^{-1/2} G_j' ,  \nonumber\\
z_i^{norm} &=  z_i  & u_i^{norm} &= m^{-1/2} u_i , \nonumber \\
v_i^{norm}&= m^{-1/2}  v_i & \alpha_i^{norm} &=m^{-1/2} \alpha_i  + 2(m^{-1/2}-1) . \nonumber
\end{align}

\section{Jet Classification with Unbalanced Optimal Transport} 
\label{sec:classification}

We now demonstrate the practical relevance of linearized, unbalanced optimal transport to collider physics by applying the pluOT framework to the task of boosted jet tagging. Discrimination between boosted jets stemming from the decay of heavy particles and QCD backgrounds composed of quark and gluon jets is a key component of many analyses at the LHC. Here we focus exclusively on distinguishing boosted $W$ boson jets from QCD backgrounds for ready comparison to previous studies applying optimal transport techniques to boosted jet tagging \cite{Komiske:2019fks, PhysRevD.102.116019}, though the same analysis can be easily extended to other pairwise tagging tasks.

Compared to previous transport-only methods \cite{PhysRevD.102.116019}, unbalanced optimal transport now allows for the creation and destruction of energy in comparing distributions. For the Hellinger-Kantorovich distance, the relative importance of mass creation/destruction over mass transportation is controlled by the intrinsic length scale parameter $\kappa$, as explained in Sec.~\ref{sec:unbalanced} and more extensively in \cite{2021arXiv210208807C}. Insofar as collider physics applications typically feature one or more additional length scales (such as the jet clustering radius $R$), a key question is how HK-based classification depends on the value of $\kappa$ relative to other length scales. To this end, we consider values of $\kappa$ ranging over several orders of magnitude, i.e., $\kappa \in [0.01, 100]$. 

\begin{figure*}
	\centering
	\begin{tabular}{lr}
		\includegraphics[width=\textwidth]{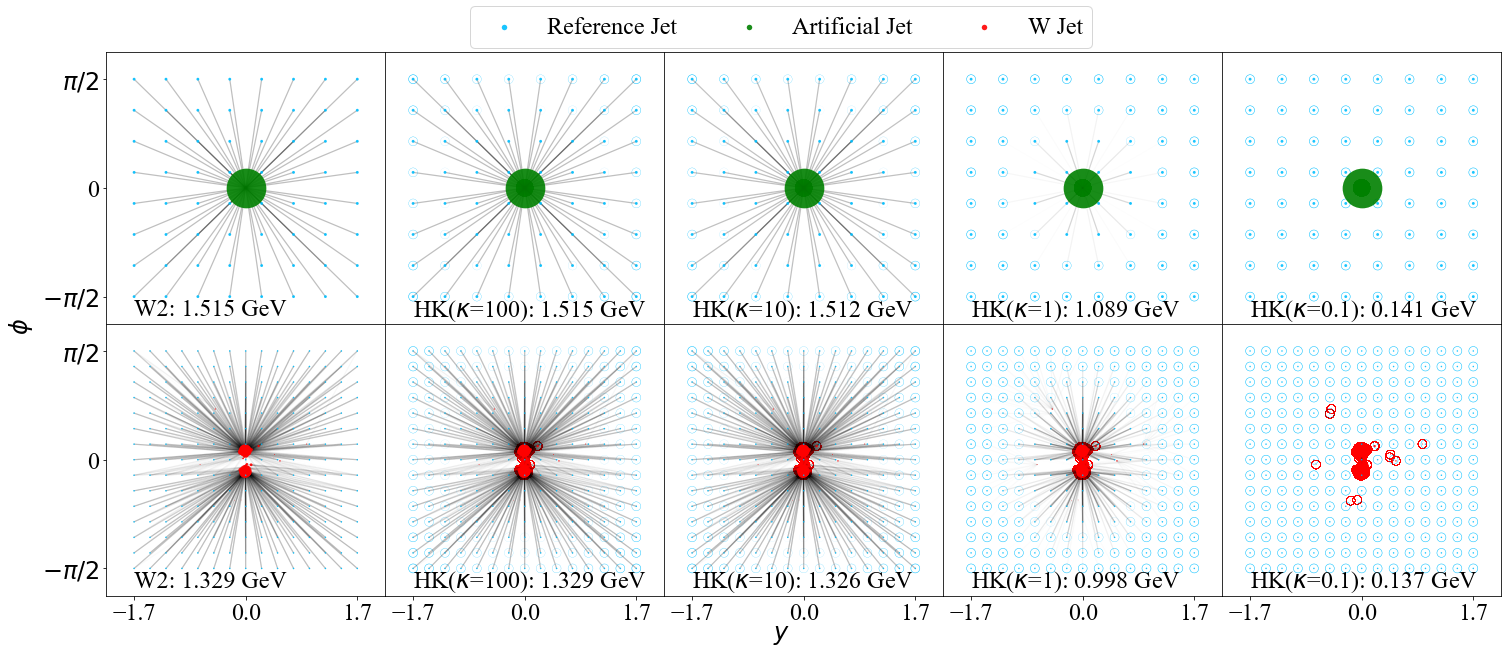}
	\end{tabular}
	\caption{Optimal transport plans from uniform reference measures  to jets. In the top row, the reference measure (blue) consists of   $8\times8$ particles and is transported to an artificial jet (green), composed of a single particle at the origin. In the second row, the uniform reference measure (blue) consists of  $15\times15$ particles and is  transported to a simulated $W$ jet (red). The columns correspond to different choices of OT metric: $W_2$ and HK$^\kappa$ with $\kappa = 100, 10, 1, 0.1$ (from left to right). The size of the filled dots indicates the amount of $p_T$ at that point. The darkness of the lines indicates how much $p_T$ is moved from one particle to another. For HK$^\kappa$, the thickness of the circles around the points represents how much $p_T$ is destroyed for that particular particle. Also shown at the bottom of the plots are the total OT distances between the jets, which are similar for $\kappa=+\infty, 100, 10$, the transport regime.}
	\label{fig:EventDisplay}
\end{figure*}

To develop intuition for the behavior of the linearized Hellinger-Kantorovich as a function of $\kappa$, in \Cref{fig:EventDisplay} we show the optimal transport plans corresponding to either the $W_2$ or HK$^{\kappa}$ metrics for various values of $\kappa$; see \cref{Wpdef,HKKantorovich}. With an eye towards linearization, the OT plans are calculated between sample jets and uniform reference measures. In the first row, OT plans are constructed between an artificial jet composed of a single particle and a reference measure consisting of an $8\times8$ grid of particles. In the second row, OT plans are constructed between a simulated boosted $W$ jet and a $15\times15$-particle uniform measure.

For the single-particle artificial jet, both the mass transportation and mass creation/destruction are symmetric with respect to the origin. Less and less mass is created or destroyed as $\kappa$ is increased, and the HK distance approaches the value of the $W_2$ distance; see the first three columns in \Cref{fig:EventDisplay}. On the other hand, a small $\kappa$ essentially reduces the HK distance to the Hellinger distance, which corresponds to the ordinary Euclidean difference between rescaled images, where no mass is being transported; see \cref{Hellinger}.  The regime of intermediate $\kappa$ is perhaps most interesting, as contributions from both mass transportation and creation/destruction are important.

To study the performance of the linearized Hellinger-Kantorovich distance in boosted jet classification, we consider simulated data consisting of 200k boosted $W$ jets and QCD jets, generated as in \cite{PhysRevD.102.116019}. Proton-proton collision events at $\sqrt{s} = 14$ TeV are simulated in \textsc{madgraph} 2.9.2 \cite{Alwall_2014} with $W$ bosons being pair produced, gluons generated via $q\overline{q}\to Z\to\nu\overline{\nu}g$, and quarks via $qg\to Z\to\nu\overline{\nu}q$. The particles are then hadronized and decayed in \textsc{Pythia} 8.302 \cite{Sj_strand_2015}, where default tuning and showering parameters are used. Afterwards, we cluster the events into jets using \textsc{FastJet} 3.3.4 \cite{Cacciari_2012} with anti-$k_T$ algorithm (jet radius $R=1$) where at most two jets are kept with $|y| \leq 1.7$ and $|\phi| \leq \frac{\pi}{2}$. Before calculating their linearized $W_2$ or HK embedding, we boost and rotate the jets to center the jet 4-momentum and vertically align the principal component of the constituent $p_T$ flow in the $y-\phi$ plane using the \texttt{EnergyFlow} package \cite{Komiske_2018,Komiske_2019,Komiske:2019asc,Komiske:2019fks,Komiske:2019jim}. The pre-processing procedure removes artificial differences in the energy flows of the jets.

Once the Euclidean embedding for each jet is acquired via pluOT (see  \cref{Eucemb}), we employ simple machine learning algorithms such as k-Nearest Neighbors (kNN) and Support Vector Machine (SVM) to classify the jets; see  Section IV in \cite{PhysRevD.102.116019} for a detailed discussion of the models. Given that our main interest here is the optimal transport distance itself, we limit ourselves mostly to kNN and do not consider other potentially more powerful machine learning models or deep neural networks, though their use can be easily implemented and incorporated into the present framework. 

We consider classification of boosted jets in datasets consisting of either 10k or 200k $W$ and QCD jets. For datasets with 10k jets, we use 5000 jets to train kNN and SVM, 2500 for validation in order to pick the best model hyperparameter(s), and the remaining 2500 jets as the test dataset to obtain the model performance. We try $k \in [10, 100]$ with an increment of 10 for kNN, and $C, \gamma \in [10^{-2}, 10^5]$ for SVM where only powers of 10 are considered. These ranges are chosen by experience so as to ensure the coverage of the optimal value of the hyperparameter(s) but as little of anything else as possible. When dealing with the full 200k dataset, we use 150k jets to train the models and 50k to evaluate the performance, where the model hyperparameter(s) are already picked by smaller runs with the 10k datasets.

We compare the tagging performance of kNN based on pluOT to that of N-subjettiness $\tau_N$, a popular jet substructure observable designed to capture the prongedness of a jet \cite{Thaler_2011,Thaler_2012}. Since $W$ jets typically have two prongs and QCD jets are more diffuse and single-pronged, the N-subjettiness ratio $\tau_{21} = \tau_2 / \tau_1$ is particularly well suited to the task at hand.  Here $\tau_N$ is determined using the \texttt{Nsubjettiness} plug-in package in \textsc{FastJet} \cite{Thaler_2011,Thaler_2012}.

Another benchmark for gauging the performance of pluOT is the pairwise EMD distance matrix \cite{Komiske:2019fks} coupled with the same machine learning models. We test the EMD both on normalized jets, where the jets are first rescaled to have $p_T = 1$, as well as on unnormalized jets. The ability to   compare both normalized and unnormalized jets is implemented by a built-in function in the \texttt{EnergyFlow} package \cite{Komiske_2018,Komiske_2019,Komiske:2019asc,Komiske:2019fks,Komiske:2019jim}, with the parameters $R, \beta = 1$ and the normalization parameter \texttt{norm} set respectively to \texttt{True} or \texttt{False}.

In a similar manner, the pluOT framework also presents us with two options to calculate the Euclidean embedding. One way is to compute the unbalanced OT distances directly between  jets with different total $p_T$, as   in \cref{HKKantorovich}. Alternatively, we can again  normalize the jets so that each has $p_T = 1$ and then compute the unbalanced HK$^{\kappa}$ distance between the normalized jets. We emphasize that, even when two jets have equal total $p_T$, as in the case of balanced OT, the HK distance still allows for local mass to be created and destroyed. 
In fact, the normalized and unnormalized approaches are  related by simple scaling transformations, see \cref{normalizedHKcoords}. For this reason, in practice we begin by computing the Euclidean embedding of normalized jets and then invert \cref{normalizedHKcoords} to recover the embedding of the unnormalized jets. Hereafter, we abbreviate the distances calculated on normalized jets with a subscript of \textit{N} and those obtained for unnormalized jets with subscript \textit{unN}.     

Previous studies of jet classification based on OT have been relatively insensitive to differences in total $p_T$ among different jets in the sample, typically considering events drawn from narrow (50 GeV) $p_T$ bins. Indeed, in \cite{PhysRevD.102.116019} it was observed that classification based on balanced optimal transport distances between normalized jets drawn from a 50 GeV $p_T$ bin modestly outperformed unbalanced optimal transport distances using the EMD. To better assess the effects of unbalanced samples, we explore events drawn from a broader range of total $p_T$, extending from $[500, 550]$ GeV in our previous study \cite{PhysRevD.102.116019} to $[500, 1500]$ GeV. This is achieved by stacking 20 datasets, each containing 10k jets with a $p_T$ bin of 50 GeV, i.e., $p_T \in [500, 550]$ GeV for the first dataset, $p_T \in [550, 600]$ GeV for the second, and so forth. In this way, in addition to the 20 datasets each with 50 GeV $p_T$ bin width, we have a combined dataset of 200k jets in which the total jet $p_T$ is approximately uniformly distributed between 500 and 1500 GeV. We examine the pluOT framework on the classification of jets with widely different total $p_T$ in Sec.~\ref{sec:classificationA}. 

When working with linearized OT, a reference measure should be chosen in advance against which the OT distance of simulated jets are computed. Loosely speaking, the reference measure is the point on the manifold of events that defines the tangent plane for linearized OT. Our default is a uniform jet with a total $p_T = 750$ GeV and $15 \times 15 = 225$ constituent particles, whose $p_T$ is distributed uniformly on the $y-\phi$ rectangle $[-1.7, 1.7] \times [- \pi/2, \pi/2]$. Such a reference measure has about the same number of particles as in a typical $W$ or QCD jet in our sample of simulated events. Other uniform references considered in this study have $4 \times 4 = 16$, $8 \times 8 = 64$, $30 \times 30 = 900$, and $60 \times 60 = 3600$ constituent particles. The inter-particle spacing $l$ of these reference measures differs widely, ranging from roughly 0.05 to 0.85. This defines yet another length scale in addition to the HK scale parameter $\kappa$, the jet clustering radius $R$, and the characteristic angular separation of the partonic decay products of a boosted particle of mass $m$, $\propto m / p_T$. We study the effect of the reference spacing $l$ on the tagging performance in Sec.~\ref{sec:classificationB} and summarize the interplay of the various scales in Sec.~\ref{sec:classificationC}.

\subsection{$p_T$ Range} \label{sec:classificationA}

In this subsection, we study the performance of classification based on unbalanced optimal transport as a function of the $p_T$ range of the simulated events, comparing the tagging performance of $W$ vs.~QCD jets whose total $p_T \in [500, 550]$ GeV or $[500, 1500]$ GeV. The three OT distances examined are: 1) the EMD distance on normalized jets ($\textit{EMD}_\textit{N}$) and unnormalized jets ($\textit{EMD}_\textit{unN}$); 2) the balanced $W_2$ distance on normalized jets; and 3) the HK$^{\kappa}$ distance on both normalized and unnormalized jets (denoted as $\textit{HK}_\textit{N}$ or $\textit{HK}_\textit{unN}$). The N-subjettiness ratio $\tau_{21}$ is also computed for each jet where classification using $\tau_{21}$ serves as a benchmark.

For the HK$^{\kappa}$ distance, we consider the $\kappa$ values $+ \infty$, 100, 10, 1, 0.5, 0.4, 0.3, 0.2, 0.1, 0.07, 0.05, 0.03, 0.01, with $\kappa = + \infty$ denoting the $W_2$ distance. Here the reference measure is taken to be a uniform jet with a total $p_T = 750$ GeV and $15 \times 15 = 225$ particles. Since it is impossible to calculate and store the entire distance matrix for 200k jets using the EMD approach with reasonable computational resources, we only compute EMD distances on the 10k datasets, whereas the linear $W_2$ and HK$^{\kappa}$ embedding can be calculated efficiently for the full 200k datasets.

\begin{figure*}
    \centering
    \includegraphics[width=\textwidth]{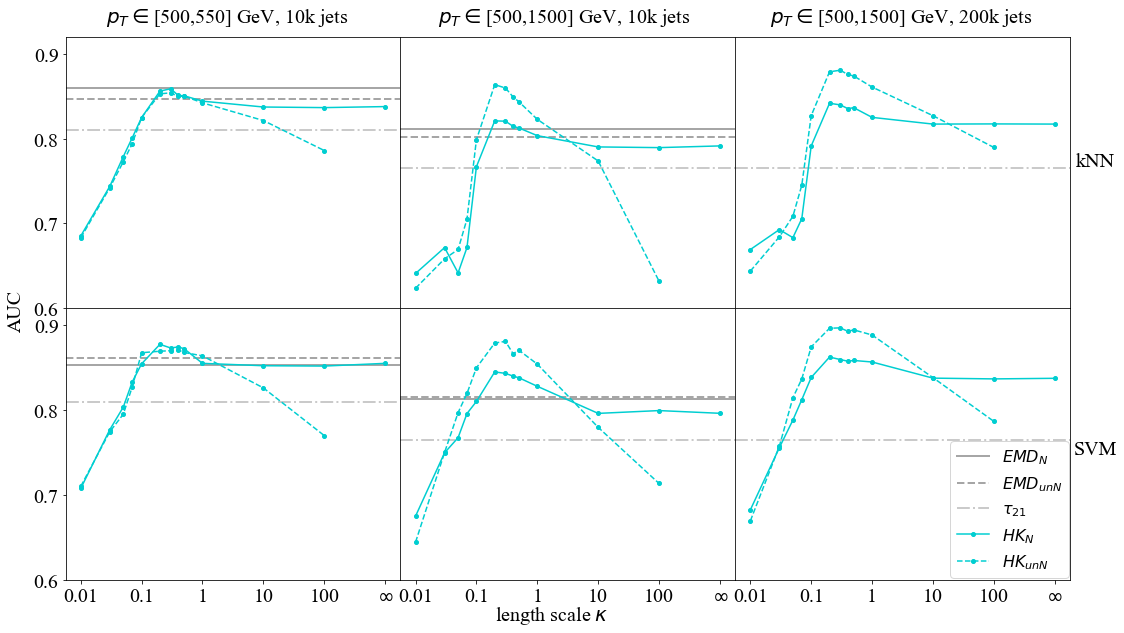}
    \caption{AUC scores for classifying $W$ vs.~QCD jets using kNN and SVM models coupled to linear $W_2$/HK$^{\kappa}$ embedding with $\kappa \in [0.01, + \infty]$. Jet $p_T$ ranges from 500 to 550 (1500) GeV in the first (second, third) column. The datasets for Column 1 and Column 2 (Column 3) have 10k (200k) jets. Solid (dashed) blue lines show the results calculated on normalized (unnormalized) jets; horizontal grey solid (dashed) lines use the EMD metrics on normalized (unnormalized) jets; and grey dash-dotted lines give the performance using $\tau_{21}$ as the discriminator.}
    \label{fig:AUC_pT500andpT500to1500}
\end{figure*}

\Cref{fig:AUC_pT500andpT500to1500} shows the tagging performance in terms of the AUC score, a number in $[0, 1]$ where 1 indicates a perfect classifier and 0.5 corresponds to random guessing. A more detailed table including the true positive rate (TPR), the false positive rate (FPR), and the optimal hyperparameters is presented in Appendix~\ref{app:Tables}, where results from other tasks in Sec.~\ref{sec:classificationB} are also included. A discussion of the general trends of the tagging performance not specific to the present task is deferred to Sec.~\ref{sec:classificationC}. 

As can be seen, for jets drawn from a 50 GeV-wide $p_T$ bin (column 1), classification performance on either normalized or unnormalized jets is almost indistinguishable for linearized HK$^{\kappa}$ distances with small $\kappa$ values ($\kappa \leq$ 1). The EMD approach also produces similar AUC scores regardless of whether or not the jets are normalized, with kNN slightly preferring the normalized approach and SVM favoring the unnormalized version. The percentage differences in the AUC are within $1.5\%$, consistent with statistical fluctuations. Such behavior is to be expected since normalization should not make a big difference when the total $p_T$ difference among jets is small. Additionally, the tagging performance of the LOT approximation, including $W_2$ and HK$^{\kappa}$ (with the exception of HK$^{\kappa}$\_unN for large $\kappa$) approaches the same (or better, in the case of SVM) level of accuracy of the EMD method, with far less computational expense. 

The effect of normalization becomes significant when the $p_T$ bin width is broadened. For jets with $p_T \in [500, 1500]$ GeV (10k for column 2 and 200k for column 3), the HK distance with $\kappa$ in its optimal range calculated directly on the unnormalized jets (dashed blue lines) gives superior performance to the normalized jets (solid blue lines), whether we use kNN or SVM as the coupled model. The increase in AUC reaches about $5\%$ at its peak when $\kappa\sim0.2$. There the AUC from the HK distance, whether normalized or not, is noticeably higher than when using the EMD distance.

Interestingly, such performance gain is not observed in the EMD approach. Here it makes no notable difference whether we use $\textit{EMD}_\textit{N}$ (solid grey line) or $\textit{EMD}_\textit{unN}$ (dashed grey line). This implies that though the difference in total jet $p_T$ has potential discriminating power, not all approaches to unbalanced optimal transport take advantage of it. A simple difference term like $|p_T(\text{jet 1}) - p_T(\text{jet 2})|$, as included in the original EMD formulation, does not lead to improved discrimination for samples drawn from a larger $p_T$ range. In contrast, unbalanced HK$^{\kappa}$, especially $\textit{HK}_\textit{unN}$, appears to take better advantage of this information by allowing local mass to be created and destroyed in addition to being transported.  

Note that, while the original formulation of the EMD in the particle physics literature considered a fixed scale parameter $R = \frac{\kappa}{2} \geq \max_{ij} d_{ij}/2$ coinciding with the jet clustering radius, one could perform a similar analysis by using the more general partial transport distance, investigating how different choices of $R = \frac{\kappa}{2}$ lead to different amounts of creation and destruction and, potentially, improved AUC in certain regimes. However, due to the fact that such metrics lack a Riemannian structure amenable to linearization, the analysis of finding the optimal parameter $R = 2 \kappa$ would be extremely computationally intensive.

\subsection{Reference Measures} \label{sec:classificationB}

In the pluOT framework, we are in principle free to pick any reasonable measure as our reference jet. Ideally, the choice of a reference measure should not exert too large an impact on the calculated linear $W_2$/HK embedding and the downstream tagging performance. Since the reference measure is associated with its own scale, the inter-particle spacing $l$, it is natural to consider the interplay between $l$ and the HK scale parameter $\kappa$ in determining what constitutes a reasonable measure in practice. 

\begin{figure*}
    \centering
    \includegraphics[width=\textwidth]{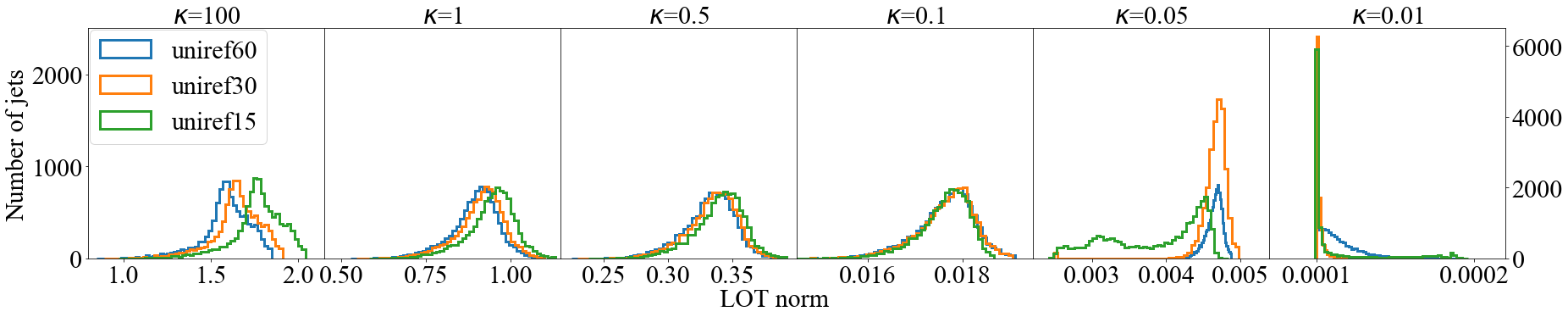}
    \caption{Distribution of the LOT norm, i.e., the distance from each jet's LOT coordinate to the origin, for the HK distance with $\kappa=100,1,0.5,0.1,0.05,0.01$. The uniform reference measures used include uniref15 with $15\times15$ particles, uniref30, and uniref60. The y coordinate of the rightmost plot follows the scale on the right, while the other plots follow the scale on the left.}
    \label{fig:LOT_hist}
\end{figure*}

To this end, we examine five uniform reference jets consisting of $4\times4$, $8\times8$, $15\times15$ (the default), $30\times30$, and $60\times60$ particles, respectively denoted as uniref4, uniref8, uniref15, uniref30 and uniref60. In \Cref{fig:LOT_hist}, we show the distribution of the Euclidean norms of the LOT coordinates of 10k jets ($p_T \in [500, 550]$ GeV) with HK$^{\kappa}$ using uniref15, uniref30, and uniref60.\footnote{As we will see, the uniref4 and uniref8 reference measures are too coarse to capture the relevant structure of the jets for any value of $\kappa$, and the distribution of Euclidean norms for these measures are correspondingly omitted from \Cref{fig:LOT_hist}.} As $\kappa$ is decreased from large values $\kappa \sim 100$ the distribution of the norms using the HK$^{\kappa}$ distance becomes more and more similar for different reference measures. The closest agreement occurs for $\kappa \sim 0.1$, which we will see later is the $\kappa$ value that gives the optimal tagging performance. As $\kappa$ is decreased below $\kappa \sim 0.1$ and we enter a scaled Euclidean image difference regime, the discrepancy of the norms using different reference measures becomes noticeable. We will see that this instability with respect to the chosen reference measure translates to deterioration of the tagging performance for small $\kappa$ values. 
 
\begin{figure*}
    \centering
    \includegraphics[width=\textwidth]{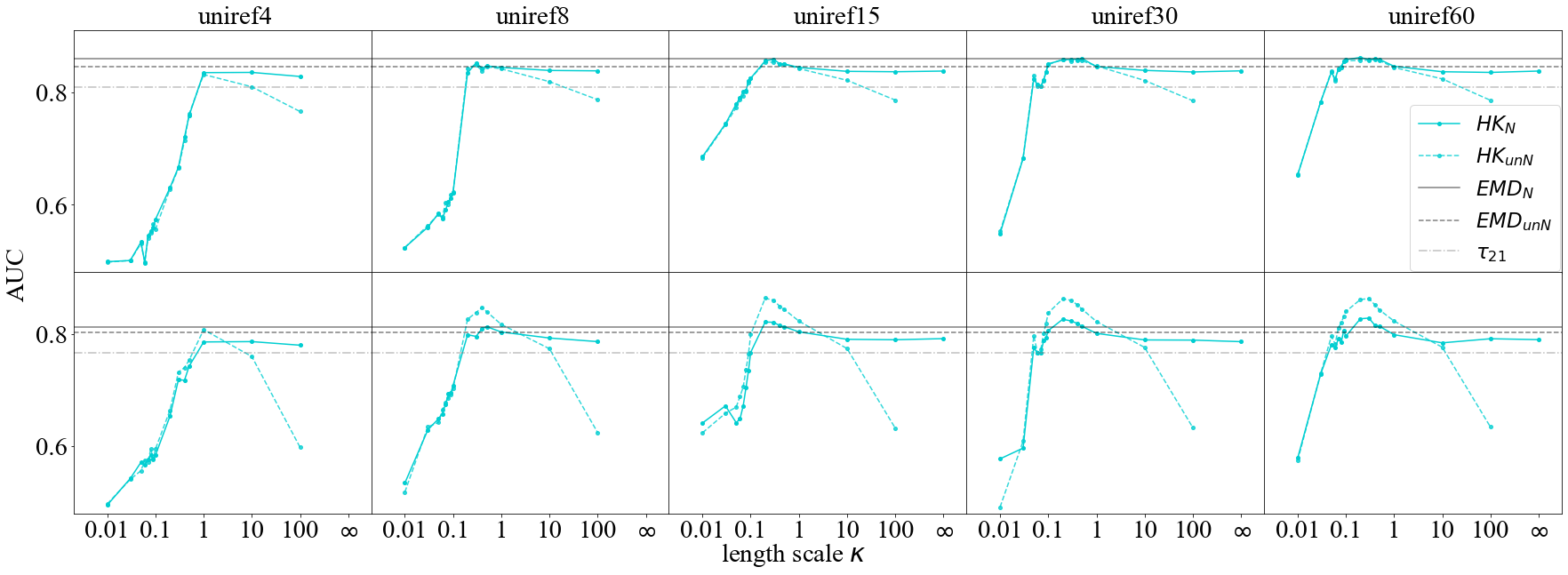}
    \caption{AUC scores for classifying 10k $W$ vs.~QCD jets using different reference measures, with uniref4, uniref8, uniref15, uniref30, and uniref60 (from left to right). The machine learning model used here is kNN. Jet $p_T$ is in between 500 and 550 (1500) GeV in the first (second) row. Solid (dashed) blue lines show the results calculated on normalized (unnormalized) jets for $W_2$/HK distance; horizontal grey solid (dashed) lines use the EMD metrics on normalized (unnormalized) jets; and grey dash-dotted lines give the tagging performance of $\tau_{21}$.\label{fig:AUC_DiffRef_full_kNN_EMD}}
\end{figure*}

\Cref{fig:AUC_DiffRef_full_kNN_EMD} shows the tagging performance on 10k jets with total $p_T \in [500, 550]$ GeV (first row) and $[500, 1500]$ GeV (second row) using $\textit{EMD}_\textit{N}$, $\textit{EMD}_\textit{unN}$; $\textit{HK}_\textit{N}$, $\textit{HK}_\textit{unN}$; and the N-subjettiness ratio $\tau_{21}$. Tagging performance is plotted in terms of AUC as a function of $\kappa$ for the HK distances. Apart from similar behaviors already discussed in Sec.~\ref{sec:classificationA}, we observe here that the peak tagging performance is roughly the same for all reference measures except uniref4, which does not attain tagging performance comparable to any EMD distance for any value of $\kappa$. Although uniref8 yields tagging performance comparable to $\textit{EMD}_\textit{unN}$ for jets with total $p_T \in [500, 550]$ GeV, it does not reach the tagging performance of $\textit{EMD}_\textit{N}$. In contrast, the tagging performance of pluOT using uniref15, uniref30, and uniref60 meets or exceeds the tagging performance of the EMD distances for optimized values of $\kappa$. This suggests that the classification performance of the linear $W_2$/HK distance is rather robust to the choice of the reference for uniref15 and finer measures. Considering that the finest reference measure under consideration (uniref60) incurs a relatively high computational cost without significant improvement in tagging performance, in what follows we largely favor the default $15 \times 15$ reference jet, reserving some comparisons with uniref30 for Appendix \ref{app:Tables}.  

Table \ref{tab:BestKappaDiffRef} lists the $\kappa$ value that produces the best AUC score for each task using $\textit{HK}_\textit{N}$ and $\textit{HK}_\textit{unN}$ metrics. Ignoring uniref4, the optimal value $\kappa_{\text{best}}$ lies between 0.2 and 0.5 for all others, regardless of the inter-particle spacing $l$. No obvious relationship is observed between $l$ and $\kappa_{\text{best}}$.    

\begin{table}[htbp]
\begin{center}
\caption{Optimal $\kappa$ values and their corresponding AUC scores for kNN classification of $W$ vs. QCD jets using different reference measures.}
\begin{tabular}{ ||c|c||c|c||c|c|| }
\hline
 \multicolumn{2}{||c||}{\textbf{Jet $\bm{p_T}$ (GeV)}} & \multicolumn{2}{|c||}{\textbf{[500, 550]}} & \multicolumn{2}{|c||}{\textbf{[500, 1500]}} \\
 \hline
 \multicolumn{2}{||c||}{\textbf{Reference}} & \textbf{HK}$_\textbf{N}$ & \textbf{HK}$_\textbf{unN}$ & \textbf{HK}$_\textbf{N}$ & \textbf{HK}$_\textbf{unN}$ \\
 \hline
 \textbf{uniref4} & $\kappa_{\text{best}}$ & 10 & 1 & 10 & 1 \\
 \cline{2-6}
 \textbf{(4$\times$4)} & AUC & 0.835 & 0.832 & 0.786 & 0.807 \\
 \hline
 \textbf{uniref8} & $\kappa_{\text{best}}$ & 0.3 & 0.3 & 0.5 & 0.4 \\
 \cline{2-6}
  & AUC & 0.852 & 0.849 & 0.813 & 0.847 \\
 \hline
 \textbf{uniref15} & $\kappa_{\text{best}}$ & 0.3 & 0.3 & 0.2 & 0.2 \\
 \cline{2-6}
  & AUC & 0.859 & 0.854 & 0.821 & 0.863 \\
 \hline
 \textbf{uniref30} & $\kappa_{\text{best}}$ & 0.5 & 0.2 & 0.2 & 0.2 \\
 \cline{2-6}
  & AUC & 0.860 & 0.859 & 0.826 & 0.862 \\
 \hline
 \textbf{uniref60} & $\kappa_{\text{best}}$ & 0.2 & 0.4 & 0.3 & 0.3 \\
 \cline{2-6}
  & AUC & 0.862 & 0.858 & 0.828 & 0.863 \\
  \hline 
\end{tabular}
\label{tab:BestKappaDiffRef}
\end{center}
\end{table}

\subsection{Discussion} \label{sec:classificationC}

The tagging performance of the HK-based metrics in \Cref{fig:AUC_pT500andpT500to1500,fig:AUC_DiffRef_full_kNN_EMD} exhibits three distinct regimes as a function of $\kappa$. In the regime where mass-creation/destruction dominates ($\kappa \lesssim 0.1$), the AUC scores for both $\textit{HK}_\textit{N}$ and $\textit{HK}_\textit{unN}$ are comparable and decrease with decreasing $\kappa$. From \cref{lkdef}, we know that no mass is allowed to be moved a distance more than $\frac{\pi}{2} \kappa$. When $\kappa$ becomes so small such that $\frac{\pi}{2} \kappa < l$ (where $l$ is the inter-particle spacing of the reference jet), mass transportation is largely forbidden when computing the distance between a jet and the reference measure. Furthermore, in this regime the assumption that, for each particle $x_j'$ in the jet, there exists a particle $x_i$ in the reference measure so that $\|x_i - x_j'\| < \frac{\kappa \pi}{2}$ is often violated, causing the linearization to break down; see \cref{LHKMsection}.
While this breakdown could be avoided in a continuum formulation of the linearization \cite{2021arXiv210208807C}, one would still have to contend with the fact that, as $\kappa \to 0$, the rescaled Hellinger-Kantorovich metric converges to the Hellinger metric, in which all information on the spatial distribution of the jets is discarded and their distance is based purely on the difference between their energies at each location; see \cref{Hellinger}.
We observe this breakdown at the level of the AUCs in \Cref{fig:LOT_hist}, considering the value  $\kappa=0.01$. 
At the other end, at large $\kappa$, the tagging performance using $\textit{HK}_\textit{N}$ stabilizes for $\kappa \gtrsim 1$, whereas the AUC score deteriorates significantly using $\textit{HK}_\textit{unN}$. As $\kappa$ grows sufficiently large, it becomes increasingly expensive to create or destroy mass. Once we enter this transport-only regime, $\kappa$ no longer plays any role for $\textit{HK}_\textit{N}$. On the other hand,  whenever the total energies of the events are unequal,  $\textit{HK}_\textit{unN}$ diverges to $+\infty$ as $\kappa \to +\infty$. 

In between these two extremes, $0.1 \lesssim \kappa \lesssim 1,$ the tagging performance of both $\textit{HK}_\textit{N}$ and $\textit{HK}_\textit{unN}$ is optimized, matching or exceeding the EMD approach. In this regime both mass transportation and creation/destruction are relevant. We have not observed any strong correlations between the optimal value $\kappa_{\text{best}}$, reference spacing $l$, the jet clustering radius $R$, and the typical angular separation of boosted partonic decay products $\propto m / p_T$, and no definite conclusion can be drawn at this stage regarding the dependence of $\kappa_{\text{best}}$ on various jet length scales. We leave this question to future studies.

A major advantage of the pluOT framework over EMD is that pluOT significantly speeds up computation and requires much less storage space. On average, only about an hour is needed to calculate the LOT distance for 10k jets on a laptop, with further speedup in the HK$^{\kappa}$ embedding for smaller $\kappa$ values. In contrast, computing EMD for 10k jets takes approximately 15 hours for jets drawn from a 50 GeV-wide $p_T$ bin, and 30-40 hours for jets drawn from a 1 TeV-wide $p_T$ bin. 

To make matters worse, the EMD outputs a huge distance matrix, which for large datasets is impossible to store on a local computer or even on a moderate cluster. The matrix is then directly fed into downstream machine learning models and since there are fewer models that can directly handle the distance matrix rather than the coordinate of each jet, the EMD approach also limits the  practical choices of ML models. Of course, the model training step is now more time-efficient than pluOT since it now only needs to read in the pairwise distance from the matrix instead of computing the Euclidean distance on the fly. But the difference in time is not significant: for kNN, EMD takes seconds while pluOT takes minutes; and for SVM, EMD needs minutes whereas pluOT requires about three hours. This gain in the latter ML step is not big enough to offset the huge time gap in OT computation. 

\vspace{0.1cm}

\section{Pileup Robustness of Optimal Transport-based Classification \label{sec:pileup}}

Pileup contamination reduces the efficacy of many commonly used jet physics observables \cite{CMS:2013wea,ATLAS:2012jma,Aad:2015ina} such as jet mass and dijet mass, where in \cite{Salam:2010nqg} the impact of different levels of pileup on dijet mass is studied. This in turn motivates the invention of various pileup mitigation techniques \cite{Soyez:2018opl,Berta:2014eza,Krohn:2013lba,Bertolini:2014bba,Cacciari:2014gra,Komiske:2017ubm}. Pileup mitigation has recently been recast in the language of optimal transport \cite{Komiske:2020qhg}, but the robustness of OT-based approaches to jet classification has yet to be studied. Here we carry out a preliminary study of the effects of pileup on $W$/QCD jet classification. 

Again we use the same $W$ and QCD dataset with $p_T\in[500,550]$ GeV and jet radius $R=1$ as described in Sec. \ref{sec:classification}. Background contamination is generated in \textsc{Pythia} where the actual number of pileup events per bunch crossing follows a Poisson distribution around $\langle N_{PU}\rangle$. We consider three different pileup benchmarks with $\langle N_{PU}\rangle=20,80,140$. These pileup templates are then added to each event and \textsc{FastJet} is used to group the pileup-contaminated events into jets. We then follow the same procedure as before, applying the pluOT framework to the pileup-contaminated jets. 

Here three reference measures are included: the default $15\times15$ uniform reference; the $30\times30$ uniform reference; and a ``pileup'' reference jet picked from one of the pileup templates for each value of $\langle N_{PU}\rangle$. For example, when examining jets contaminated by pileup with $\langle N_{PU} \rangle=80$, the reference measure is taken to be another Poisson distribution with $\langle N_{PU} \rangle=80$. The motivation behind the choice of uniref30 is that since the number of particles in the reference is close to that of the jets contaminated by pileup with $\langle N_{PU} \rangle=80, 140$, uniref30 should better capture the true underlying differences between $W$ vs.~QCD jets not obscured by the superficial pileup addition.

\begin{figure*}
    \centering
    \includegraphics[width=\textwidth]{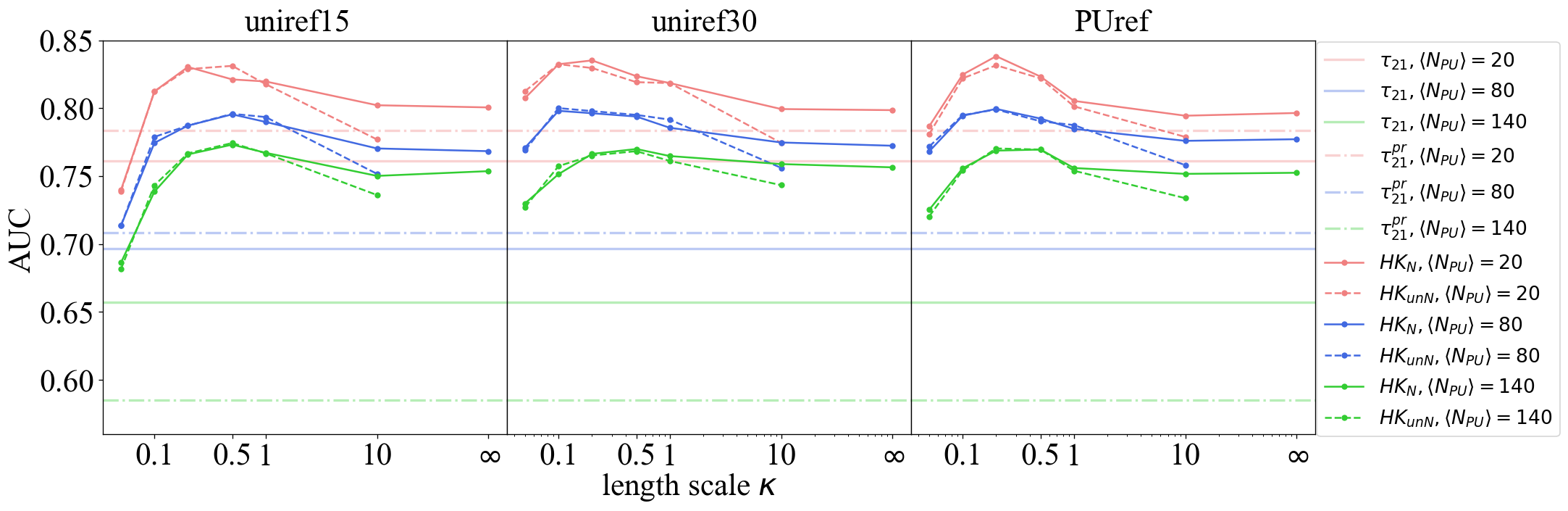}
    \caption{AUC scores for using kNN to classify 10k $W$ vs.~QCD jets with different amount of pileup where the average numbers of pileup particles in each event are $\langle N_{PU} \rangle = 20$ (red), 80 (blue), and 140 (green). From left to right, the reference measures used are the $15\times15$ uniform reference, the $30\times30$ uniform reference, and a jet drawn from the pileup template corresponding to each $N_{PU}$. As usual, solid (dashed) lines show the AUC scores using the linear HK/$W_2$ distances on normalized (unnormalized) jets, and solid horizontal lines give the tagging performance of $\tau_{21}$ on unpruned jets whereas the dash-dotted lines are the results using $\tau_{21}$ on pruned jets (denoted by $\tau_{21}^{pr}$).}
    \label{fig:AUC_JetPU}
\end{figure*}

Again, the N-subjettiness ratio $\tau_{21}$ serves as a benchmark, where $\tau_{21}$ is computed both on the datasets with and without pileup. The one without pileup is generated by pruning the contaminated datasets, accomplished in \textsc{FastJet} by a pruner that reclusters the jets with Cambridge-Aachen algorithm and removes constituent particles that are soft or at large angles with other particles \cite{Ellis:2009su,Ellis:2009me}. \Cref{fig:AUC_JetPU} displays the resulting AUC vs.~$\kappa$ curves, where we use kNN+linear $W_2$/HK distances with $\kappa = +\infty, 10, 1, 0.5, 0.2, 0.1$, and $0.05$ on both normalized and unnormalized jets. 

It is clear from the figure that comparing to $\tau_{21}$ (horizontal lines), the tagging performance of pluOT behaves rather well and does not decay significantly as pileup increases. Especially for high pileup scenarios, the AUC scores of kNN+$W_2$/HK distances on un-pruned jet samples using any of the three references are far better than the corresponding AUCs of kNN+$\tau_{21}$, where for $\langle N_{PU} \rangle=140$, $\tau_{21}$ on pruned jets behaves much worse than that on un-pruned jets, corroborating the observation in \cite{CMS:2013uea} that N-subjettiness on groomed jets is less discriminant than being computed on ungroomed jets. More studies need to be performed in order to examine in detail the influence of background contamination such as pileup on OT-based metrics, but its potential advantage over traditional methods is already clear. 

\section{Conclusion \label{sec:conclusion}}

The Energy Mover's Distance \cite{Komiske:2019fks} illustrates the many advantages to be gained by equipping the space of collider events with a metric, from unifying the panoply of collider observables to enabling the use of interpretable distance-based machine learning algorithms. These successes invite further exploration of the larger space of optimal transport distances with an eye towards collider physics. In this paper we have generalized the EMD by situating it within a  multi-parameter family of unbalanced optimal transport metrics. Among the other metrics in this family, the Hellinger-Kantorovich distance stands out as particularly suited to collider applications insofar as it (uniquely) preserves a Riemannian structure. The resulting manifold of collider events inherits various satisfying properties, including a well-defined tangent space at each point on the manifold. This enables the computationally efficient linearization of unbalanced optimal transport distances \cite{2021arXiv210208807C} in close analogy with Linearized Optimal Transport in the balanced case \cite{wang2013linear,PhysRevD.102.116019}.

Motivated by these advantages, we developed the particle linearized unbalanced Optimal Transport (pluOT) framework for computing linearized Hellinger-Kantorovich distances. To illustrate the efficacy of pluOT for collider physics, we used it as input to simple distance-based machine learning algorithms for boosted $W$/QCD jet classification and studied the performance as a function of various scale parameters. For optimized parameter choices, we found that pluOT matched or exceeded the same algorithms using EMD distances in a fraction of the computing time. Although the effects of pileup on optimal transport distances have yet to be extensively studied, we found that boosted jet classification based on pluOT also exhibited an encouraging degree of robustness against pileup contamination compared to the N-subjettiness shape observable. 

There are numerous avenues for further exploration. Within the pluOT framework itself, there is considerable room to explore the interplay between the Hellinger-Kantorovich length scale parameter, the jet clustering radius, and the scale(s) associated with the choice of reference measure. We have focused on boosted jet classification as an initial application to collider physics, but pluOT should be generally well-suited to the same array of applications as the Energy Mover's Distance. More broadly, linearization is but one of the many potential advantages of applying the Hellinger-Kantorovich distance to collider physics. The Riemannian event manifold obtained with the Hellinger-Kantorovich distance is likely to have interesting properties and may reveal further hidden structure in the space of collider events.

\begin{acknowledgments}
We thank B.~Schmitzer and M.~Thorpe for many helpful conversations and for sharing their code to compute the Hellinger-Kantorovich distance. The work of T.~Cai and N.~Craig is supported by the U.S.~Department of Energy under the grant DE-SC0011702. N.~Craig thanks LBNL and the BCTP for hospitality during the completion of this work. T.~Cai and J.~Cheng are grateful for a Worster Fellowship from the UCSB Department of Physics. The work of K. Craig is supported by NSF DMS grant 1811012 and a Hellman Faculty Fellowship. K. Craig gratefully acknowledges the support from the Simons Center for Theory of Computing, at which this work was completed. 
\end{acknowledgments}

\appendix

\section{Result Tables \label{app:Tables}}

Here we include two tables with numerical results for the classification tasks mentioned in Sections \ref{sec:classification} and \ref{sec:pileup}. Both tables consist of AUC scores for $W$ vs.~QCD jet classification using the kNN model. In Table \ref{tab:table1} the dataset has default jet radius $R=1$, and the reference measures used are $15 \times 15$ uniform measure (uniref15) and $30 \times 30$ uniform measure (uniref30). Jet $p_T$ ranges are $[500,550]$ GeV with 10k jets, $[500,1500]$ GeV with 10k jets, and $[500,1500]$ GeV with 200k jets. Also included are tagging results using EMD metrics and $\tau_{21}$.

\begin{table*}[htbp]
\caption{AUC scores for kNN classification of $W$ vs.~QCD jets.}
\begin{center}
\begin{adjustbox}{scale=0.95, center}
\begin{tabular}{|c|c|c|c|c|c|}
    \hline
    \multirow{2}{*}{ref jet} & \multirow{2}{*}{$\kappa$} & \multirow{2}{*}{Normalization} & \multicolumn{3}{|c|}{$p_T$}\\
    \cline{4-6}
    & & & $[500,550]$ GeV, 10k jets & $[500,1500]$ GeV, 10k jets & $[500,1500]$ GeV, 200k jets\\
    \hline
    \hline
    \multirow{17}{*}{uniref15} & $+\infty$ & N & 0.838 & 0.791 & 0.817\\
    \cline{2-6}
    & \multirow{2}{*}{100} & N & 0.836 & 0.789 & 0.817\\
    & & unN & 0.786 & 0.632 & 0.790\\
    \cline{2-6}
    & \multirow{2}{*}{10} & N & 0.837 & 0.790 & 0.817\\
    & & unN & 0.821 & 0.774 & 0.827\\
    \cline{2-6}
    & \multirow{2}{*}{1} & N & 0.844 & 0.803 & 0.825\\
    & & unN & 0.842 & 0.823 & 0.861\\
    \cline{2-6}
    & \multirow{2}{*}{0.5} & N & 0.850 & 0.812 & 0.836\\
    & & unN & 0.850 & 0.843 & 0.874\\
    \cline{2-6}
    & \multirow{2}{*}{0.2} & N & 0.856 & 0.821 & 0.842\\
    & & unN & 0.853 & 0.863 & 0.879\\
    \cline{2-6}
    & \multirow{2}{*}{0.1} & N & 0.825 & 0.767 & 0.791\\
    & & unN & 0.825 & 0.799 & 0.827\\
    \cline{2-6}
    & \multirow{2}{*}{0.05} & N & 0.779 & 0.642 & 0.683\\
    & & unN & 0.773 & 0.669 & 0.708\\
    \cline{2-6}
    & \multirow{2}{*}{0.01} & N & 0.685 & 0.641 & 0.669\\
    & & unN & 0.683 & 0.624 & 0.644\\
    \hline
    \hline
    \multirow{17}{*}{uniref30} & $+\infty$ & N & 0.838 & 0.786 & 0.815\\
    \cline{2-6}
    & \multirow{2}{*}{100} & N & 0.836 & 0.789 & 0.815\\
    & & unN & 0.785 & 0.633 & 0.791\\
    \cline{2-6}
    & \multirow{2}{*}{10} & N & 0.839 & 0.789 & 0.815\\
    & & unN & 0.821 & 0.776 & 0.827\\
    \cline{2-6}
    & \multirow{2}{*}{1} & N & 0.846 & 0.801 & 0.827\\
    & & unN & 0.847 & 0.822 & 0.860\\
    \cline{2-6}
    & \multirow{2}{*}{0.5} & N & 0.860 & 0.813 & 0.839\\
    & & unN & 0.856 & 0.844 & 0.874\\
    \cline{2-6}
    & \multirow{2}{*}{0.2} & N & 0.857 & 0.826 & 0.842\\
    & & unN & 0.859 & 0.862 & 0.880\\
    \cline{2-6}
    & \multirow{2}{*}{0.1} & N & 0.851 & 0.806 & 0.827\\
    & & unN & 0.849 & 0.837 & 0.861\\
    \cline{2-6}
    & \multirow{2}{*}{0.05} & N & 0.823 & 0.775 & 0.802\\
    & & unN & 0.830 & 0.797 & 0.825\\
    \cline{2-6}
    & \multirow{2}{*}{0.01} & N & 0.549 & 0.577 & 0.566\\
    & & unN & 0.552 & 0.492 & 0.567\\
    \hline
    \hline
    \multirow{2}{*}{EMD} & \multicolumn{2}{|c|}{N} & 0.859 & 0.812 & \multirow{2}{*}{N/A}\\
    & \multicolumn{2}{|c|}{unN} & 0.846 & 0.802 &\\
    \hline
    \hline
    \multicolumn{3}{|c|}{$\tau_{21}$} & 0.810 & 0.766 & 0.765\\
    \hline
\end{tabular}
\end{adjustbox}
\end{center}
\label{tab:table1}
\end{table*}

In Table \ref{tab:table2} the pileup contaminated dataset has the default jet radius $R=1$ and their $p_T$ range is $[500,550]$ GeV, with pileup levels being $\langle N_{PU}\rangle=20,80,140$. The reference measures used are uniref15, uniref30, and a jet drawn from the pileup template corresponding to each $N_{PU}$. The tagging results using $\tau_{21}$ on jets both with and without pruning are included as well.

\begin{table*}[h!]
\caption{AUC scores for kNN classification of $W$ vs.~QCD jets with different levels of pileup.}
\begin{center}
\begin{adjustbox}{scale=0.95, center}
\begin{tabular}{|c|c|c|c|c|c|}
    \hline
    ref jet & $\kappa$ & Normalization & $\langle N_{PU}\rangle=20$ & $\langle N_{PU}\rangle=80$ & $\langle N_{PU}\rangle=140$\\
    \hline
    \hline
    \multirow{13}{*}{uniref15} & $+\infty$ & N & 0.801 & 0.768 & 0.754\\
    \cline{2-6}
    & \multirow{2}{*}{10} & N & 0.802 & 0.770 & 0.750\\
    & & unN & 0.777 & 0.752 & 0.736\\
    \cline{2-6}
    & \multirow{2}{*}{1} & N & 0.820 & 0.790 & 0.767\\
    & & unN & 0.818 & 0.794 & 0.767\\
    \cline{2-6}
    & \multirow{2}{*}{0.5} & N & 0.821 & 0.796 & 0.773\\
    & & unN & 0.831 & 0.796 & 0.774\\
    \cline{2-6}
    & \multirow{2}{*}{0.2} & N & 0.830 & 0.787 & 0.766\\
    & & unN & 0.829 & 0.787 & 0.767\\
    \cline{2-6}
    & \multirow{2}{*}{0.1} & N & 0.812 & 0.775 & 0.739\\
    & & unN & 0.812 & 0.779 & 0.743\\
    \cline{2-6}
    & \multirow{2}{*}{0.05} & N & 0.740 & 0.714 & 0.686\\
    & & unN & 0.739 & 0.714 & 0.682\\
    \hline
    \hline
    \multirow{13}{*}{uniref30} & $+\infty$ & N & 0.799 & 0.772 & 0.757\\
    \cline{2-6}
    & \multirow{2}{*}{10} & N & 0.799 & 0.775 & 0.759\\
    & & unN & 0.775 & 0.756 & 0.743\\
    \cline{2-6}
    & \multirow{2}{*}{1} & N & 0.819 & 0.786 & 0.765\\
    & & unN & 0.819 & 0.792 & 0.761\\
    \cline{2-6}
    & \multirow{2}{*}{0.5} & N & 0.824 & 0.794 & 0.770\\
    & & unN & 0.819 & 0.795 & 0.768\\
    \cline{2-6}
    & \multirow{2}{*}{0.2} & N & 0.835 & 0.796 & 0.766\\
    & & unN & 0.830 & 0.798 & 0.765\\
    \cline{2-6}
    & \multirow{2}{*}{0.1} & N & 0.832 & 0.798 & 0.752\\
    & & unN & 0.832 & 0.800 & 0.757\\
    \cline{2-6}
    & \multirow{2}{*}{0.05} & N & 0.808 & 0.771 & 0.730\\
    & & unN & 0.813 & 0.769 & 0.727\\
    \hline
    \hline
    \multirow{13}{*}{PUref} & $+\infty$ & N & 0.797 & 0.777 & 0.753\\
    \cline{2-6}
    & \multirow{2}{*}{10} & N & 0.795 & 0.776 & 0.752\\
    & & unN & 0.779 & 0.758 & 0.734\\
    \cline{2-6}
    & \multirow{2}{*}{1} & N & 0.805 & 0.785 & 0.756\\
    & & unN & 0.801 & 0.788 & 0.754\\
    \cline{2-6}
    & \multirow{2}{*}{0.5} & N & 0.823 & 0.792 & 0.770\\
    & & unN & 0.822 & 0.790 & 0.770\\
    \cline{2-6}
    & \multirow{2}{*}{0.2} & N & 0.838 & 0.800 & 0.769\\
    & & unN & 0.832 & 0.799 & 0.770\\
    \cline{2-6}
    & \multirow{2}{*}{0.1} & N & 0.825 & 0.795 & 0.756\\
    & & unN & 0.822 & 0.795 & 0.754\\
    \cline{2-6}
    & \multirow{2}{*}{0.05} & N & 0.787 & 0.768 & 0.725\\
    & & unN & 0.781 & 0.772 & 0.720\\
    \hline
    \hline
    \multicolumn{3}{|c|}{$\tau_{21}$} & 0.761 & 0.697 & 0.657\\
    \hline
    \multicolumn{3}{|c|}{pruned $\tau_{21}$} & 0.784 & 0.708 & 0.585\\
    \hline
\end{tabular}
\end{adjustbox}
\end{center}
\label{tab:table2}
\end{table*}

\clearpage

\bibliography{emdbib}

\end{document}